# Frictionally decaying frontal warm-core eddies

Angelo Rubino

*Dipartimento di Scienze Ambientali, Informatica e Statistica, Università Ca'Foscari, Venice, Italy*

Sergey Dotsenko

*Marine Hydrophysical Institute, Sevastopol, Russia*

Corresponding author: Angelo Rubino, Dipartimento di Scienze Ambientali, Informatica e Statistica, Università Ca' Foscari di Venezia, Campus Scientifico di via Torino, edificio Zeta, Via Torino 155, 30175 Mestre (VE), Italia, tel.: +39 041 234 8406, email: rubino@unive.it

# ABSTRACT

The dynamics of nonstationary, nonlinear, axisymmetric, warm-core geophysical surface frontal vortices affected by Rayleigh friction is investigated semi-analytically using the nonlinear, nonstationary reduced-gravity shallow-water equations. In this frame, it is found that vortices characterized by linear distributions of their radial velocity and arbitrary structures of their section and azimuthal velocity can be described exactly by a set of nonstationary, nonlinear coupled ordinary differential equations. The first-order problem (i.e., that describing vortices characterized by a linear azimuthal velocity field and a quadratic section) consists of a system of 4 differential equations, and each further order introduces in the system three additional ordinary differential equations and two algebraic equations.

In order to illustrate the behavior of the nonstationary decaying vortices, the system's solution for the first-order and for the second-order problem is then obtained numerically using a Runge-Kutta method. The solutions demonstrate that inertial oscillations and an exponential attenuation dominate the vortex dynamics: Expansions and shallowings, contractions and deepenings alternate during an exact inertial period while the vortex decays. The dependence of the vortex dissipation rate on its initial radius is found to be non-monotonic: It is higher for small and large radii. Our analysis adds realism to previous theoretical investigations on mesoscale vortices, represents an ideal tool for testing the accuracy of numerical models in simulating nonlinear, nonstationary frictional frontal phenomena in a rotating ocean, and paves the way to further extensions of (semi-) analytical solutions of hydrodynamical geophysical problems to more arbitrary forms and more complex density stratifications.

# 1. INTRODUCTION

Geophysical frontal vortices are frequently observed in the ocean (see, e.g., McWilliams 1985; Olson 1991). Particularly, they are believed to play a fundamental role in different, important oceanic phenomena like, e.g., those related to the transfer of physical, chemical, and biological properties across frontal regions (see, e.g., Saunders 1971; Cheney et al. 1976; Armi and Zenk 1984; Joyce 1984; Olson et al. 1985; Dengler et al. 2004), to the formation and transformation of water masses (see, e.g., Gascard et al., 2002; Budeus et al., 2004), and to the downward propagation of wind generated near-inertial waves (Lee and Niiler, 1998; Zhai et al., 2007). This extraordinarily large relevance explains why, in the last decades, oceanic frontal vortices have been deeply investigated experimentally, analytically and numerically (see, e.g., Csanady, 1979; Gill, 1981; Nof, 1983; McWilliams, 1985, 1988; Rubino and Brandt 2003; Rubino et al. 2002; Rubino et al. 2009).

A prominent role in the theoretical investigation of geophysical frontal vortices has been played by the nonlinear, reduced gravity shallow water equations (Cushman-Roisin 1987; Cushman-Roisin and Merchant-Both 1995; Cushman-Roisin et al., 1985; Rubino et al. 1998; Rubino et al. 2003; Rubino et al. 2006; Rubino et al. 2009).

Although these equations do not allow for the development of baroclinic instabilities or for the radiation of energy toward the interior ocean via internal waves propagation, their use enables to explain different characteristics of observed frontal vortices (Cushman-Roisin 1987; Cushman-Roisin and Merchant-Both 1995; Cushman-Roisin et al., 1985; Rubino et al. 1998; Rubino et al. 2003; Rubino et al. 2006; Rubino et al. 2009). Moreover, the possibility of expressing analytically the evolution of a broad class of geophysical frontal vortices using these equations represents a valuable way for testing the accuracy of numerical models for non-stationary, nonlinear, frontal phenomena of geophysical relevance (Cushmann Roisin 1986; Cushmann Roisin 1987; Rubino et al. 2002; Rubino and Brandt 2003).

Fundamentally, the features emerging as analytical solutions in the reduced-gravity frame are characterized by inertial oscillations affecting both radial and tangential velocities and vortex's depth: The vortices contract and expand during an exact inertial period (Cushman-Roisin 1987; Rubino et al. 1998). Note that, in these analytical solutions, the vortex's radial velocity field is constrained to be a linear function of the vortex's radius, while the azimuthal velocity field and the vortex's section can show a more complex shape (Rubino et al. 1998; Dotsenko and Rubino 2006).

On the (unavoidably obscure) way of searching for analytical solutions of systems of unsteady, nonlinear coupled partial differential equations the fundamental step consists in their reduction to a system of ordinary differential equations. Has this step been performed, so different methods can be attempted to solve analytically the obtained system. But, even if an analytical solution cannot be found, numerical methods often allow then for an extremely accurate simulation of the involved dynamics, as the originally multidimensional problem has been reduced to a monodimensional one.

In the case of frontal mesoscale vortices of the ocean, one has also to note that, among the different scales characterizing their temporal variability (from the superinertial ones affecting the swirl velocity of elongated elliptical anticyclones to the exactly inertial one typical of the circular pulson and to the super inertial -subinertial modes emerging in warm-core eddies on a β-plane, see, e.g., Cushmann-Roisin et al. 1985; Rubino et al. 2009) that associated with their frictional decay plays a particular role, because it contributes to determine the fate of these mesoscale features: In the case of warm-core rings, for instance, their remarkable longevity enables disintegration in coastal areas and/or re-absorption by the parent current to fundamentally contribute to their dissipation/variability (Fierl and Mied 1985; Tomosada 1986; Sachihiko et al 2011).

In the present investigation, the dynamics of nonstationary, nonlinear, axisymmetric, warm-core geophysical surface frontal vortices affected by Rayleigh friction is analyzed semi-analytically using the nonlinear, nonstationary reduced-gravity shallow-water equations. Our analysis adds realism to previous theoretical investigations on mesoscale vortices, represents an ideal tool for testing the accuracy of numerical models in simulating nonlinear, nonstationary frictional frontal phenomena in a rotating ocean, and paves the way to further extensions of (semi-) analytical solutions of hydrodynamical geophysical problems to more arbitrary forms and more complex density stratifications.

The paper is organized as follows: In the next section the mathematical model is illustrated and its reduction to a system of ordinary differential equations is presented. Numerical solutions obtained using a Runge-Kutta methods are discussed in section 3. Finally, the obtained results are discussed und conclusions drawn in section 4

## 2. THE MATEMATICAL MODEL

In the frame of the non-stationary, nonlinear reduced-gravity equations on an $f$-plane we consider the frictional (Rayleigh) dynamics of a circular, frontal warm-core eddy: A lens of light water of density $\rho$ lies on the top of a heavier, infinitely deep quiescent ocean of density $\rho^*$ (Fig. 1). Assuming circular symmetry, the motion of the active layer can be expressed as follow in cylindrical coordinates $(r, z, \phi)$:

$$\frac{\partial u}{\partial t} + u\frac{\partial u}{\partial r} - \frac{v^2}{r} - fv = -g'\frac{\partial h}{\partial r} - su, \qquad (1)$$

$$\frac{\partial v}{\partial t} + u\frac{\partial v}{\partial r} + \frac{uv}{r} + fu = -sv, \qquad (2)$$

$$\frac{\partial h}{\partial t} + \frac{1}{r}\frac{\partial (ruh)}{\partial r} = 0, \qquad (3)$$

In the expressions above, $t$ is the time, $\{u; v\}(r, t)$ are, respectively, the radial and the azimuthal projections of the horizontal velocity, $h(r, t)$ is the vortex thickness, $g' = g(1-\rho/\rho^*)$ the reduced gravity (where $g$ represents the acceleration of gravity), while $f$ is the (constant) Coriolis parameter, and

$s$ the (constant) friction coefficient.

Any solution of the system (1) – (3) has to satisfy the condition

$$h(r_0, t) = 0 \tag{4}$$

on the movable surface circular boundary of the vortex (its surface rim) located at $r = r_0$ (see Fig. 1).

**Reduction of model to a system of ordinary differential equations.**

Let us now assume that velocity fields and vortex thickness be characterized by the following horizontal structures:

$$u = A(t)r, \quad v = \sum_{i=1}^{N} B_i(t) r^{2i-1}, \quad h = \sum_{j=0}^{2N-1} C_j(t) r^{2j}, \tag{5}$$

where $A$, $B_i$ and $C_j$ are functions of time only, and $N \geq 1$. Note that the condition that the thickness $h$ of the vortex be positive and satisfy (5) implies

$$\sum_{j=1}^{2N-1} C_j(0) > 0.$$

A substitution of (5) into (1) – (3) yields the following system of nonstationary, nonlinear, coupled, ordinary differential equations (ODEs) and algebraic equations in the unknown functions $A$, $B_i$ and $C_j$:

$$\delta_{i1}\frac{dA}{dt} + \delta_{i1}A^2 - \sum_{j=1}^{i} B_j B_{i-j+1} - fB_1 + 2igC_i + sA_i = 0 \quad (i = 1,...,2N-1), \tag{6}$$

$$\frac{dB_i}{dt} + 2iAB_i + \delta_{i1}fA + sB_i = 0 \quad (i = 1,...,N), \tag{7}$$

$$\frac{dC_i}{dt} + 2(i+1)AC_i = 0 \quad (i = 0,...,2N-1), \tag{8}$$

where $\delta_{i1}$ is the Kronecker delta, and $B_i = 0$ for $i > N$. The initial conditions for the coefficients $A$, $B_i$ and $C_i$ are:

$$A(0) = A_0, \quad B_i(0) = B_{i0}, \quad C_i(0) = C_{i0}. \tag{9}$$

**First-order reduction ($N = 1$).**

In order to discuss fundamental characteristics of the dynamics of frictional vortices described above, in the following we will focus on their first-order and second-order expressions.

Let us consider the first-order expression of the vortex fields expressed in (5) – (7):

$$u = A(t)r, \, v = B_1(t)r, \, h = C_0(t) + C_1(t)r^2, \tag{10}$$

with $A$, $B_1$, $C_0$ and $C_1$ unknown functions of time only. The requirement that the vortex has positive thickness $h$ and the condition (4) yield:

$$C_0(t) > 0, \, C_1(t) < 0. \tag{11}$$

Substituting the expressions (9) into (1) – (3) and equating the coefficients of equal powers in $r$ leads to a system of four ODEs in the variables $A$, $B_1$, $C_0$ and $C_1$:

$$\frac{dA}{dt} + A^2 - B_1^2 - fB_1 + 2g'C_1 + sA = 0, \tag{12}$$

$$\frac{dB_1}{dt} + 2AB_1 + fA + sB_1 = 0, \tag{13}$$

$$\frac{dC_0}{dt} + 2AC_0 = 0, \tag{14}$$

$$\frac{dC_1}{dt} + 4AC_1 = 0. \tag{15}$$

The system of equations (12) – (15) must be supplemented with the following initial conditions:

$A(0) = A_0$,  $B_1(0) = B_{10}$,  $C_0(0) = C_{00} > 0$,  $C_1(0) = C_{10} < 0$. (16)

**Second-order reduction (N=2).** Let us now pass to the second-order expression of the vortex fields:

$u = A(t)r$,  $v = B_1(t)r + B_2(t)r^3$,  $h = C_0(t) + C_1(t)r^2 + C_2(t)r^4 + C_3(t)r^6$, (17)

with $A$, $B_{1,2}$ and $C_{0,1,2,3}$ unknown functions of time only. Substituting the expressions (17) into (1) – (3) and equating the coefficients of equal powers in $r$ leads to a system of seven ODEs and two algebraic equations in the variables $A$, $B_{1,2}$, $C_{0,1,2,3}$:

$$\frac{dA}{dt} + A^2 - B_1^2 - fB_1 + 2g'C_1 + sA = 0,$$ (18)

$$2B_1B_2 + fB_2 - 4g'C_2 = 0,$$ (19)

$$B_2^2 - 6g'C_3 = 0,$$ (20)

$$\frac{dB_1}{dt} + 2AB_1 + fA + sB_2 = 0,$$ (21)

$$\frac{dB_2}{dt} + 4AB_2 + sB_2 = 0,$$ (22)

$$\frac{dC_0}{dt} + 2AC_0 = 0,$$ (23)

$$\frac{dC_1}{dt} + 4AC_1 = 0,$$ (24)

$$\frac{dC_2}{dt} + 6AC_2 = 0,$$ (25)

$$\frac{dC_3}{dt} + 8AC_3 = 0.$$ (26)

**Analytical solution of the first-order problem without dissipation.** To illustrate the general characteristics of the dynamics of the inviscid (*s*=0) solutions of (6) – (8) (see, e.g., Cushmann-Roisin, 1985; Rubino et al. 1998) we will shortly summarize the properties of the first-order analytical solution. In this case, indeed, one finds:

$$A = \frac{1}{2}\gamma f \Psi \cos\Phi, \quad B_1 = -\frac{1}{2}f + l\Psi, \quad C_0 = c\Psi, \quad C_1 = -\frac{c}{R_0^2}\Psi^2, \tag{27}$$

where $R_0$ is the initial vortex radius,

$$\Psi = \frac{1}{1+\gamma\sin\Phi}, \quad \Phi = ft + \varphi, \text{ and } l = \frac{1}{2}f\sqrt{1-\gamma^2 - \frac{8g'c}{f^2 R_0^2}}.$$

Here $\gamma \in [0, 1)$, $\varphi$, and $c>0$ are constants. The solution (27) describes a pulsating vortex: Expansions and shallowings, contractions and deepenings alternate during an exact inertial period (Cushmann-Roisin 1985; Rubino et al. 1998). Note that the pulsation amplitude is determined by the value of $\gamma$. In the numerical experiments of the following section this solution will be used to define the (ageostrophic) initial conditions for calculating of the vortex oscillations in the frictional case.

The kinetic, potential, and total energy, $E_{kin}$, $E_{pot}$, and $E_{tot}$ of the vortex, whose structure is described by (6) – (8) can be expressed as:

$$E_{pot} = \frac{1}{2}\rho g' \iint_S h^2 dxdy, \quad E_{kin} = \frac{1}{2}\rho \iint_S h(u^2 + v^2)dxdy,$$

where $S$ corresponds to the time-dependent vortex surface.

## 3. NUMERICAL SOLUTIONS

**First-order solutions.** The system of equations (12) – (15) with the appropriate initial conditions (16) was solved numerically using a Runge-Kutta method of the fourth-order.

In order to select realistic values for the solution we refer to the detailed statistics of the Gulf Stream rings observed between 1974 and 1983 which can be found in Broun et al. (1986).

An appropriate value for the friction coefficient $s$ in Eqs. (1), (2) can be selected to reflect observed lifetimes of geophysical oceanic vortices. Assuming an exponential decay of the vortex velocity field, $t_{life}$ will represent the vortex *e*-folding time, and hence $s = 1/t_{life}$. Typical observed values for Gulf Stream rings are (Broun et al. 1986): initial radius $R_0 = 75$ km, $t_{life} = 130$ days, latitude = 38°N, and initial depth $h_0 = 500$ m. Therefore, the corresponding friction coefficient $s = 1/t_{life} = 8,903 \times 10^{-8}$ s$^{-1}$. In our evaluations we set moreover $g' = 0,01$ m·s$^{-2}$ for the density difference between vortex and ambient water, and $\varphi = 0$.

The temporal evolution of a first-order frictional vortex is shown in Fig. 2: in particular, depicted are the radial velocity $U = A(t)r_0(t)$ and the azimuthal velocity $V = B_1(t)r_0(t)$ at the (temporally varying) vortex rim $r_0(t)$, as well as the maximum vortex thickness $H = C_0(t)$ at its center. Exact inertial oscillations clearly emerge: As in the inviscid case, expansions and shallowings, contractions and deepenings alternate during an exact inertial period (Cushmann-Roisin 1985; Rubino et al. 1998), while both velocity amplitudes and maximum thickness decrease and radius increases as time elapses, due to the action of the Rayleigh friction.

Frictional decay is also evident in the evolution of the vortex integral energy characteristics (Fig. 3). As in the inviscid case, vortex kinetic and potential energy show exact inertial anti-phase oscillations, however, they also decay in time together with the vortex total energy which monotonously decreases.

In Fig. 4 is depicted the vortex total energy decay as a function of different vortex frictional coefficients (or, equivalently, as a function of different vortex longevities). Larger frictions produce faster energy decays, the latter decreasing with increasing coefficients. Larger oscillations (expressed by the magnitude of the parameter $\gamma$) lead to faster energy dissipation (Fig. 5). In Fig. 6 the temporal evolution of the total energy is depicted as a function of the initial vortex radius. The rate of dissipation is largest for initially small and large features, and smaller for intermediate ones.

**Higher-order solutions.** As an example of the frictional vortex evolution for a high-order pulson we now present results referring to the second-order problem, which is described by the system (18) – (26). Like in the previous case, the solutions have been computed numerically using a fourth-order Runge-

11Kutta algorithm.

The evolution of the vortex velocity fields and radius closely resembles that elucidated for the fist-order solution (Fig. 7). As time elapses, the amplitudes of the oscillations (which are still exactly inertial) decrease and the radius increases as time elapses, due to the action of the Rayleigh friction.

Associated to this behavior we note a more complex evolution of the different fields. As indicated previously, in the inviscid as well as in the frictional solutions the radial velocity $u$ has to be a linear function of radius for each order and frictional coefficient in order for the problem to be reduced to a system of ordinary differential equations. The influence of friction, hence, cannot alter the shape of the radial velocity horizontal distribution: Instead, it leads to a monotonic decrease of the radial velocity amplitude with increasing values of the frictional parameter $s$. In the second-order problem considered, the inviscid tangential velocity $v$ used as initial condition for the frictional simulations largely deviates from the linear one characterizing the first-order problem. In this case (not shown), it reaches its maximum well within the vortex body and then decreases toward the periphery. Note that such kind of distributions has been found as being the one typically emerging in the first evolutionary stage of an impulsive vortex generation obtained in tank experiments devoted at investigating geophysical frontal vortices (see Rubino and Brandt 2003). Friction induces a tendency toward a decreasing amplitude and nonlinearity of the tangential velocity distribution. The region of maximum values experiences a shift toward the vortex periphery as the friction coefficient $s$ increases. Accordingly, the deviation in the form of the vortex section from a parabolic one also decreases as the friction increases.

## 4. DISCUSSION AND CONCLUSIONS

In the present paper we have analyzed aspects of the dynamics of nonstationary, nonlinear, axisymmetric, warm-core geophysical surface frontal vortices affected by Rayleigh friction in the frame of the

12nonlinear, nonstationary reduced-gravity shallow-water equations. In this frame, we have shown that, in the case of circular features characterized by radial velocities which are linear functions of the vortex radius, it is possible to reduce the problem to a set of ordinary differential equations like in the case of the pulson described by Rubino et al. (1998). This step is the fundamental one in the (unavoidably obscure) search of exact analytical solutions of complex problems expressed in terms of nonstationary, nonlinear, coupled partial differential equations depending on space and time. In the present case, however, it seems that obtaining exact analytical solutions is not a straightforward task. Hence, we concentrated on numerical solutions using a fourth-order Runge-Kutta method. These solutions, however, can be believed to represent an excellent approximation to the exact ones, as the problem has been reduced to a monodimensional one.

In order to illustrate the behavior of the decaying vortices, the system's solutions for the first-order and for the second-order problem have been considered. For both cases, inertial oscillations and an exponential attenuation dominate the vortex dynamics: Expansions and shallowings, contractions and deepenings alternate during an exact inertial period while the vortex decays. The dependence of the vortex dissipation rate on its initial radius is found to be non-monotonic: It is higher for small and large radii. Our analysis adds realism to previous theoretical investigations on mesoscale vortices. It also represents an ideal tool for testing the accuracy of three-dimensional numerical models in simulating realistic geophysical problems, as it refers to nonlinear, nonstationary frictional frontal dynamics in a rotating ocean. Moreover, it paves the way to further extensions of (semi-) analytical solutions of hydrodynamical geophysical problems to more arbitrary forms and more complex density stratifications.

**FIGURE CAPTIONS**

FIGURE 1: Radial section of the considered vortices.

FIGURE 2: Inertial oscillations and attenuation of radial (a) and azimuthal (b) components of the horizontal velocity, maximum depth (c) and radius (d) of the vortex. The initial radius of vortex is $R_0 = 75$ km, the amplitude oscillation parameter is $\gamma = 0,2$. T represents the inertial period (T = 19,49 h)

FIGURE 3: Temporal evolution of the kinetic ($E_{kin}$), potential ($E_{pot}$), and total ($E_{tot}$) energy of the vortex. The initial radius of the vortex is $R_0 = 75$ km, the amplitude oscillation parameter is $\gamma = 0,2$

FIGURE 4: Temporal evolution of the vortex total energy ($E_{tot}$) for different durations vortex longevities: *1*: $t_{life} = 60$ days; *2*: $t_{life} = 120$ days; *3*: $t_{life} = 180$ days. The initial vortex radius is $R_0 = 75$ km, the amplitude oscillation parameter is $\gamma = 0,2$

FIGURE 5: Temporal evolution of the vortex total energy ($E_{tot}$) for different values of the amplitude oscillation parameter $\gamma$: *1*: $\gamma = 0,1$; *2*: $\gamma = 0,2$; *3*: $\gamma = 0,3$. The initial vortex radius is $R_0 = 75$ km, the friction coefficient is $s = 8,903 \cdot 10^{-8}$ s$^{-1}$

FIGURE 6: Temporal evolution of the vortex total energy ($E_{tot}$) for different values of the initial vortex radius: *1*: $R_0 = 75$ km; *2*: $R_0 = 150$ km; *3*: $R_0 = 225$ km; *4*: $R_0 = 300$ km. The amplitude oscillation parameter is $\gamma = 0,2$, the friction coefficient is $s = 8,903 \cdot 10^{-8}$ s$^{-1}$

FIG. 7. Inertial attenuating oscillations of radial (a) and azimuthal (b) components of the horizontal velocity, maximum depth (c) and radius (d) of a second-order vortex solution. The vortex is characterized initially by a radius $R_0 = 75$ km, a thickness $h_0 = 500$ m, and an amplitude oscillation parameter $\gamma = 0{,}2$. The friction coefficient corresponds to a $t_{life} = 130$ days, i.e., $s = 1/\, t_{life} = 8{,}903 \times 10^{-8}$ s$^{-1}$

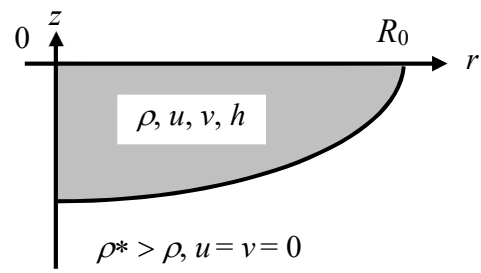

FIGURE 1: Radial section of the considered vortices

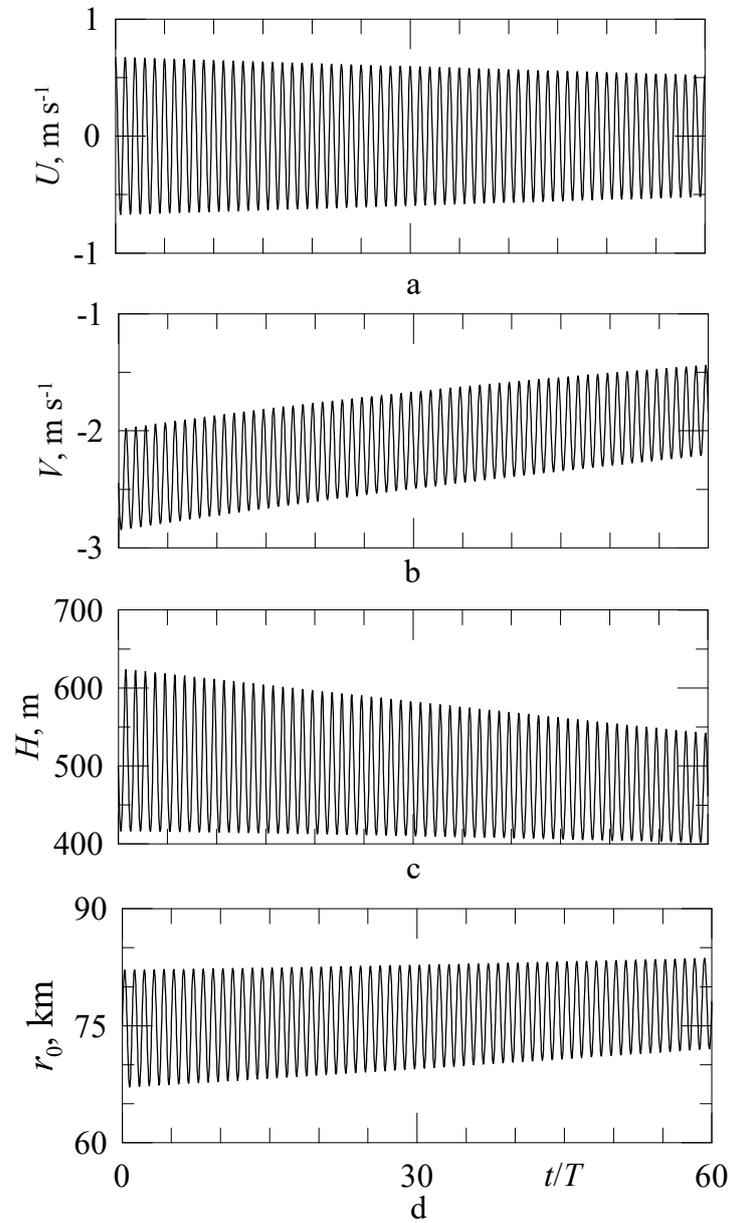

FIGURE 2: Inertial oscillations and attenuation of radial (a) and azimuthal (b) components of the horizontal velocity, maximum depth (c) and radius (d) of the vortex. The initial radius of vortex is $R_0 = 75$ km, the amplitude oscillation parameter is $\gamma = 0,2$. T represents the inertial period (T = 19,49 h)

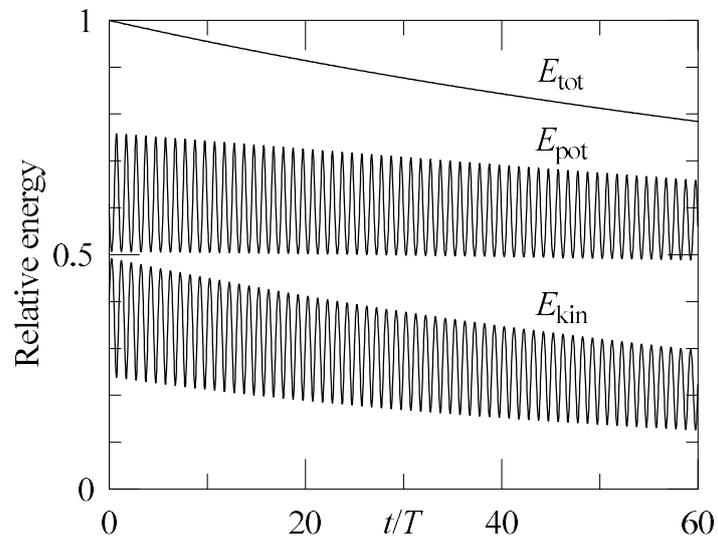

FIGURE 3: Temporal evolution of the kinetic ($E_{kin}$), potential ($E_{pot}$), and total ($E_{tot}$) energy of the vortex. The initial radius of the vortex is $R_0 = 75$ km, the amplitude oscillation parameter is $\gamma = 0,2$

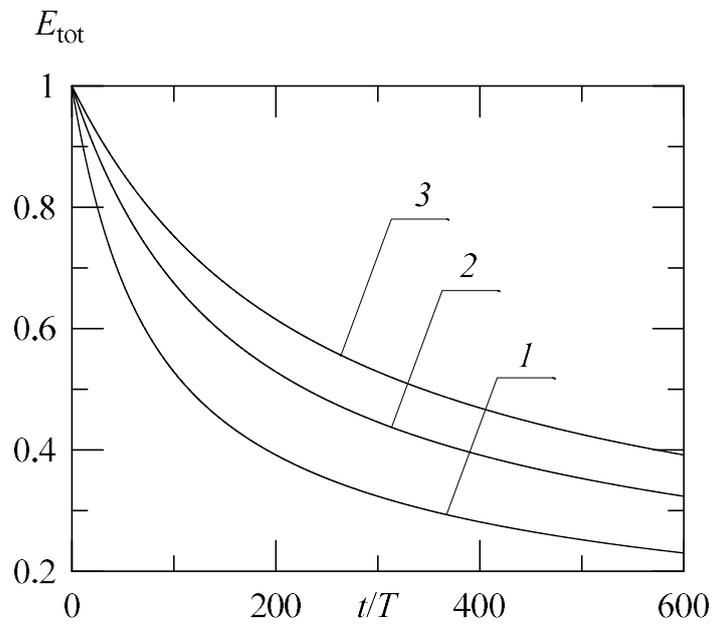

FIGURE 4: Temporal evolution of the vortex total energy ($E_{tot}$) for different durations vortex longevities: *1*: $t_{life}$ = 60 days; *2*: $t_{life}$ = 120 days; *3*: $t_{life}$ = 180 days. The initial vortex radius is $R_0$ = 75 km, the amplitude oscillation parameter is $\gamma$ = 0,2

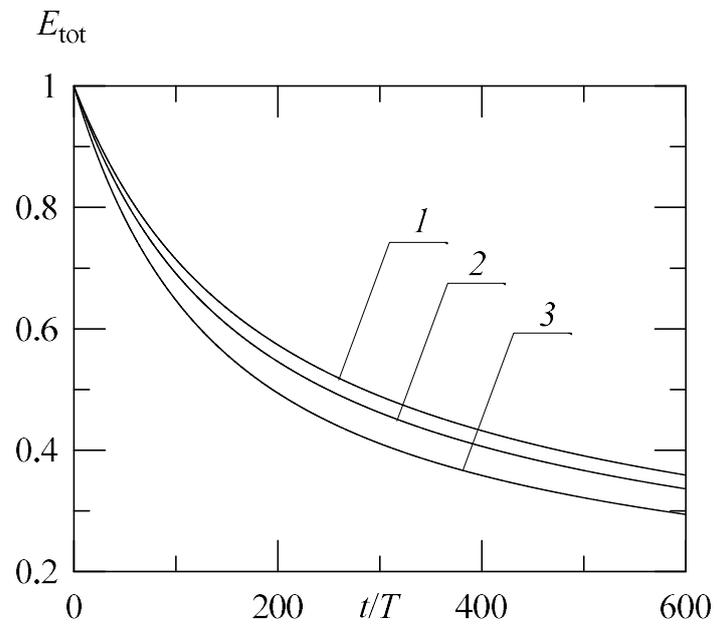

FIGURE 5: Temporal evolution of the vortex total energy ($E_{tot}$) for different values of the amplitude oscillation parameter $\gamma$: *1*: $\gamma = 0{,}1$; *2*: $\gamma = 0{,}2$; *3*: $\gamma = 0{,}3$. The initial vortex radius is $R_0 = 75$ km, the friction coefficient is $s = 8{,}903 \; 10^{-8}$ s$^{-1}$

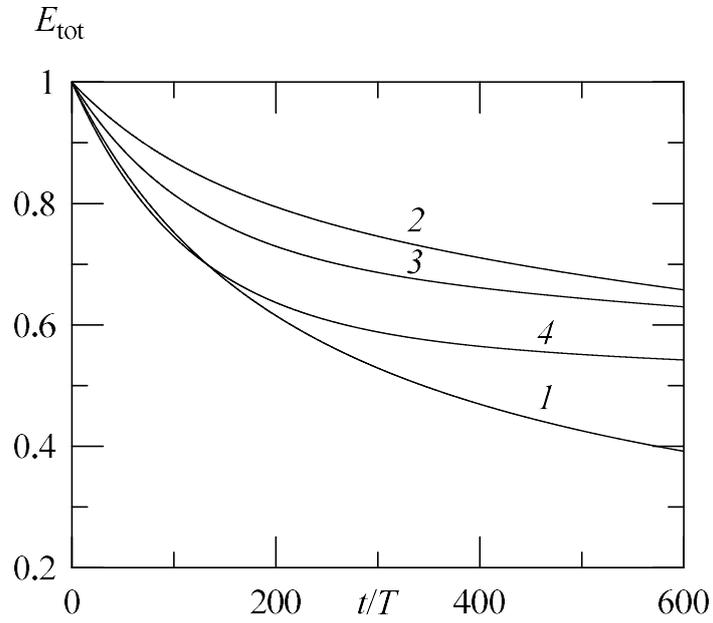

FIGURE 6: Temporal evolution of the vortex total energy ($E_{tot}$) for different values of the initial vortex radius: *1*: $R_0 = 75$ km; *2*: $R_0 = 150$ km; *3*: $R_0 = 225$ km; *4*: $R_0 = 300$ km. The amplitude oscillation parameter is $\gamma = 0{,}2$, the friction coefficient is $s = 8{,}903 \cdot 10^{-8}$ s$^{-1}$

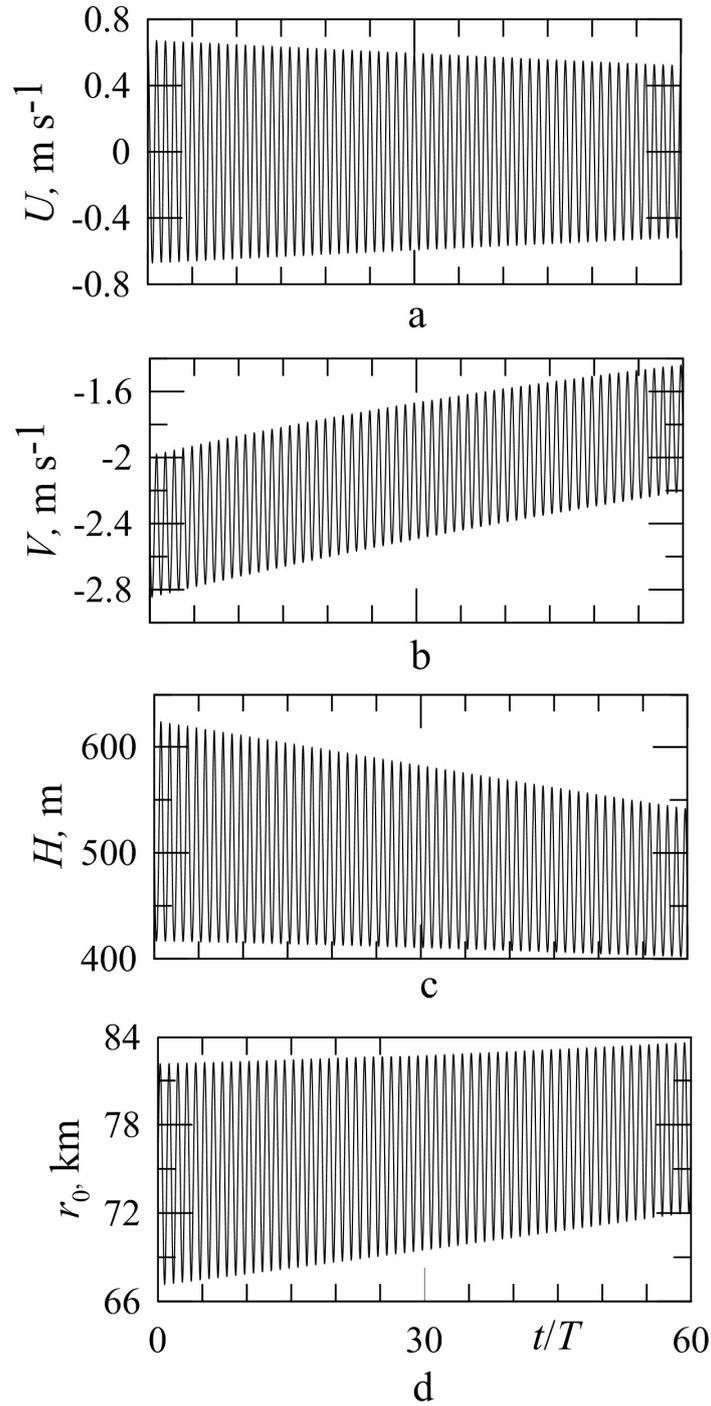

FIG. 7. Inertial attenuating oscillations of radial (a) and azimuthal (b) components of the horizontal velocity, maximum depth (c) and radius (d) of a second-order vortex solution. The vortex is characterized initially by a radius $R_0 = 75$ km, a thickness $h_0 = 500$ m, and an amplitude oscillation parameter $\gamma = 0{,}2$. The friction coefficient corresponds to a $t_{life} = 130$ days, i.e., $s = 1/\,t_{life} = 8{,}903 \times 10^{-8}$ s$^{-1}$